\documentclass{ws-procs9x6}            
\usepackage[mathscr,scaled=1.15]{urwchancal}
\DeclareFontFamily{OT1}{pzc}{}
\DeclareFontShape{OT1}{pzc}{m}{it}%
{<-> s * [1.15] pzcmi7t}{}
\DeclareMathAlphabet{\mathpzc}{OT1}{pzc}{m}{it}

\begin{document}
\title{
$\,$\\[-20ex]\hspace*{-8cm}{\textmd{\small{{Preprint no}. NJU-INP 011/19}}}\\[10ex]
Two photon transition form factors of neutral pseudoscalar mesons}

\author{Minghui Ding$^*$ and Daniele Binosi}
\address{European Centre for Theoretical Studies in Nuclear Physics
and Related Areas (ECT$^\ast$) and Fondazione Bruno Kessler\\ Villa Tambosi, Strada delle Tabarelle 286, I-38123 Villazzano (TN) Italy\\
$^*$E-mail: mding@ectstar.eu}
\vspace{-.2cm}
\author{Kh\'epani Raya and Lei Chang}
\address{School of Physics, Nankai University, Tianjin 300071, China}
\vspace{-.2cm}

\author{Craig D.~Roberts}
\address{School of Physics, Nanjing University, Nanjing, Jiangsu 210093, China\\
Institute for Nonperturbative Physics, Nanjing University, Nanjing, Jiangsu 210093, China}

\vspace{-.3cm}

\begin{abstract}
Starting from the axial vector Ward-Green-Takahashi identity, and the necessity of preserving this identity when attempting to calculate the properties describing pseudoscalar mesons, we describe a Bethe-Salpeter kernel which takes into account the non-Abelian anomaly and therefore is suitable to describe the flavor mixing states $\eta,\eta^\prime$. We then report results for the two photon transition form factors of $\eta,\eta^\prime$, and compare them with existing experimental data.

\end{abstract}


\bodymatter

\section{Ward-Green-Takahashi identity}\label{aba:sec1}
Ward's identity\cite{Ward:1950xp} and its generalization by Green\cite{Green:1953te} and Takahashi\cite{Takahashi:1957xn}, ought to be preserved by any approach that is willing to capture QCD's fundamental features. In its axial vector formulation this identity reads
\begin{align}
\nonumber P^\mu \Gamma_{5\mu}^a(k;P) & = {\cal S}^{-1}(k_+) i \gamma_5 {\cal F}^a 
+ i \gamma_5 {\cal F}^a {\cal S}^{-1}(k_-)\\
& - 2 i {\cal M}^{ab}\Gamma_5^b(k;P)  - {\cal A}^a(k;P),
\label{avwti}	
\end{align}
where: $P$ ($k$) is the total (relative) momentum between the amputated quark legs; $\{{\cal F}^a | \, a=0,\ldots,N_f^2-1\}$ are the generators of $U(N_f)$ in the fundamental representation, orthonormalized according to tr${\cal F}^a {\cal F}^b= \frac{1}{2}\delta^{ab}$; ${\cal M}^{ab} = {\rm tr}_F \left[ \{ {\cal F}^a , {\cal M^{\rm bm}} \} {\cal F}^b \right]$, with ${\cal M}^{\rm bm} = {\rm diag}[m_u^{\rm bm},m_d^{\rm bm},m_s^{\rm bm},\ldots]$ the current-quark bare masses matrix; ${\cal S}=\,$diag$[S_u,S_d,S_s,\ldots]$ the dressed-quark propagator matrix, with nonzero entries determined by the corresponding Dyson-Schwinger equation; $\Gamma_{5\mu}^a$ and $\Gamma_{5}^a$ the axial-vector and pseudoscalar vertex respectively, both satisfying the corresponding inhomogeneous Bethe-Salpeter equations; and, finally, ${\cal A}^a$ in an expression of the non-Abelian axial anomaly and can be defined through the topological charge density operator. Notice that only ${\cal A}^{a=0}$ is nonzero: therefore, flavor-nonsinglet neutral pseudoscalar mesons satisfy the standard identity, while flavor-singlet state satisfies the anomalous one~\cite{Bhagwat:2007ha}. 

One of the far-reaching consequences of the Ward-Green-Takahashi (WGT) identity~(\ref{avwti}) is a mass formula for pseudoscalar mesons which is valid irrespectively from the magnitude of the current-quark masses. It reads
\begin{equation}
\label{newmass}
m_{M}^2 f^a_{M} = 2\,{\cal M}^{ab} \rho_{M}^b + \delta^{a0}n_{M},
\end{equation} 
with: $m_{M}$ the meson mass; $f^a_{M}$ the decay constant; $\rho^b_{M}$ the pseudoscalar projection of the meson's Bethe-Salpeter wave function onto the origin in configuration space; and $n_{M}$ a quantity derived from topological charge density operator. This one single equation has a wide range of effects on Goldstone modes, excited states as well as system(s) described by the non-Abelian axial anomaly\cite{Maris:1997hd,Holl:2004un,Bhagwat:2007ha}; in particular, it shows that:
\begin{itemlist}
\item The Nambu-Goldstone pion is a consequence of Dynamical Chiral Symmetry Breaking (DCSB).
\item In the chiral limit, the decay constants of the pion's radial excitations are zero.
\item $\eta^\prime$ is split from the octet pseudoscalars by an amount that depends on QCD's topological susceptibility.
\end{itemlist}

\section{Kernels for the bound-state equations}

Yet another consequence of Eq.~(\ref{avwti}) is that it leads to a symmetry-preserving truncation scheme, the simplest of which is the Rainbow-Ladder (RL) approximation. RL truncation is known to be accurate for ground-state light-quark vector- and isospin-nonzero-pseudoscalar-mesons \cite{Horn:2016rip, Eichmann:2016yit, Fischer:2018sdj, Burkert:2019bhp}, and has been used in a recent study on pion parton distribution functions \cite{Ding:2019lwe}. Implemented in the Bethe-Salpeter and gap equations, RL truncation is expressed via the following kernels, respectively:
\begin{subequations}
\label{KDinteraction}
\begin{eqnarray}
[\mathscr{K}_N(k,q;P)]_{\iota_1\iota_2}^{\iota_1^\prime \iota_2'} & =&-
\tfrac{4}{3}{\mathpzc G}_{\mu\nu}(t) [i\gamma_\mu]_{\iota_1\iota_1'} [i\gamma_\nu]_{\iota_2^\prime\iota_2},\\
 {\mathpzc G}_{\mu\nu}(t=k-q) & =& \tilde{\mathpzc G}(t^2) T_{\mu\nu}(t),
\end{eqnarray}
\end{subequations}
where: $t^2 T_{\mu\nu}= t^2 \delta_{\mu\nu} - t_\mu t_\nu$; and $\tilde{\mathpzc G}(t^2)$ is the interaction consistent with that determined in studies of QCD's gauge sector\cite{Binosi:2014aea}.
 
However, RL truncation fails for the $\eta$- and $\eta^\prime$-mesons as they do not produce vertices that satisfy the anomalous axial-vector WGT identity. Indeed, whilst the RL kernel is constrained by a large body of successful phenomenology, the form of the anomalous kernel is unknown. Considering the structure of the non-Abelian anomaly, it readily becomes apparent that no related contribution to the Bethe-Salpeter kernel can contain external quark or antiquark lines which are simply connected to the internal lines: purely gluonic configurations must mediate.  Moreover, no finite sum of diagrams can be sufficient\cite{Bhagwat:2007ha}. On general grounds, its contribution to the Bethe-Salpeter equation for pseudoscalar mesons must take the following form\cite{Ding:2018xwy}:
\begin{align}
[{\mathpzc K}_A(k,q;P)]_{\iota_1\iota_2}^{\iota_1^\prime \iota_2^\prime} & =-\sum_{i=1}^4 \sum_{{\mathsf f}=l,s} {\mathpzc a}_i^{\mathsf f}(k,q;P)\nonumber \\
&  \times  [{\cal F}^{\mathsf f} {\cal D}_i(q;P)]_{\iota_2^\prime \iota_1^\prime} [{\cal F}^{\mathsf f} {\cal D}_i(k;P)]_{\iota_1 \iota_2},
\end{align}
where: $\{{\cal D}_i(k;P)|i=1,\ldots,4\}$ constitute a Dirac basis of the pseudoscalar meson Bethe-Salpeter wave function; and $\{{\mathpzc a}_i^{\mathsf f}(k,q;P)|i=1,\ldots,4; {\mathsf f} =l,s\}$ are scalar functions. 
An educated Ansatz for ${\mathpzc K}_A$ had been put forward in Ref.~\citenum{Ding:2018xwy}, which being essentially nonperturbative in nature, provided material support at infrared momenta. In this way the corresponding Bethe-Salpeter kernel becomes the sum of two terms: ${\mathpzc K} = {\mathpzc K}_N + {\mathpzc K}_A$, and with the bound-state kernel thus defined, results for the $\gamma^\ast \gamma \to \eta, \eta^\prime$ transition form factors have been obtained.

\section{Two-photon transition form factors}

The two-photon transition form factor provides a window for exploring the internal structure of neutral pseudoscalar meson\cite{Raya:2015gva,Raya:2016yuj,Chen:2016bpj}. The parameters in the gap and Bethe-Salpeter equations can be fixed by requiring that the solutions of the coupled-channels bound-state problems reproduce the empirical $\eta, \eta^\prime$ masses and the four phenomenologically-determined values of the light- and strange-quark $\eta$, $\eta^\prime$ decay constants\cite{Ding:2018xwy}. We find \mbox{$m_\eta=0.56$} GeV and $m_{\eta^\prime}=0.96$ GeV, to be compared with the experimental values\cite{Tanabashi:2018oca} $0.55$ GeV and $0.96$ GeV, respectively; other observables like decay constants, mixing angle and two-photon decay widths~are~reported~in~\mbox{Ref.\citenum{Ding:2018xwy}}.

\begin{figure}[t]
\begin{center}
\includegraphics[width=2.2in]{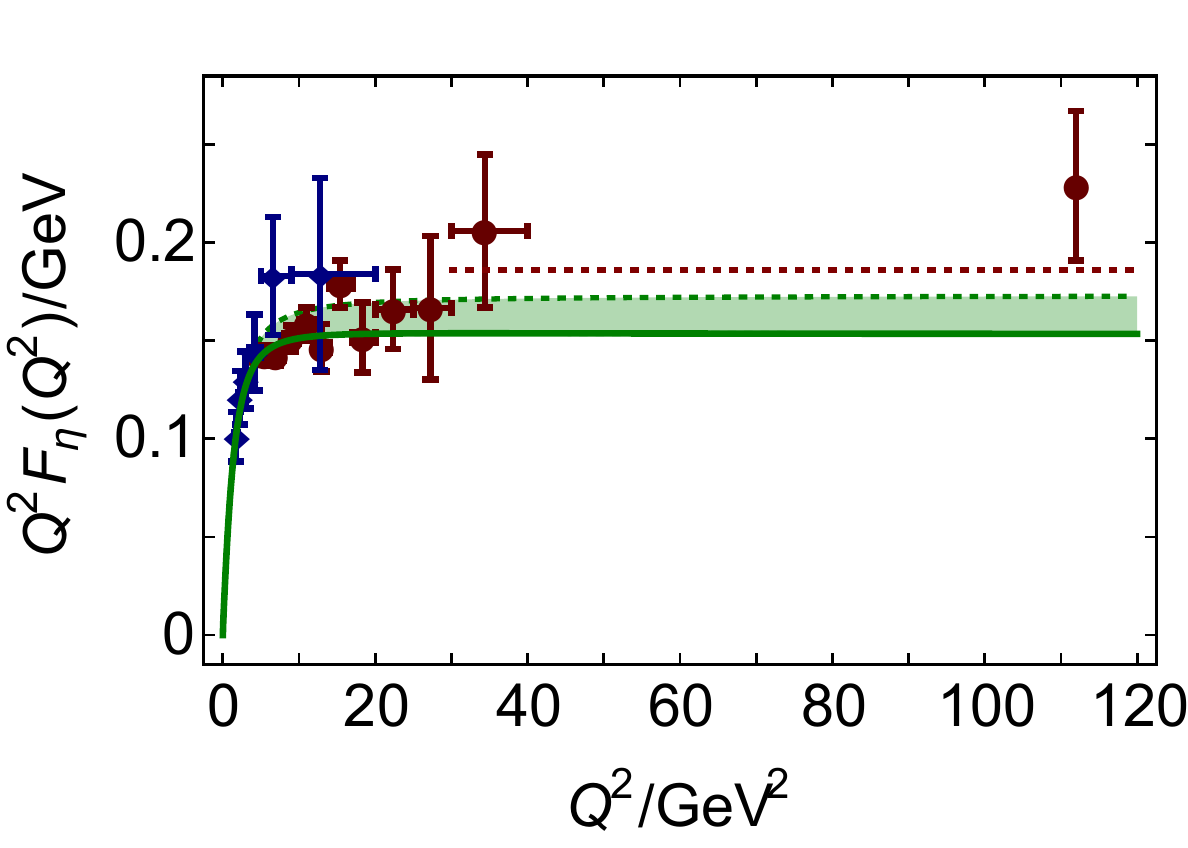}
\includegraphics[width=2.2in]{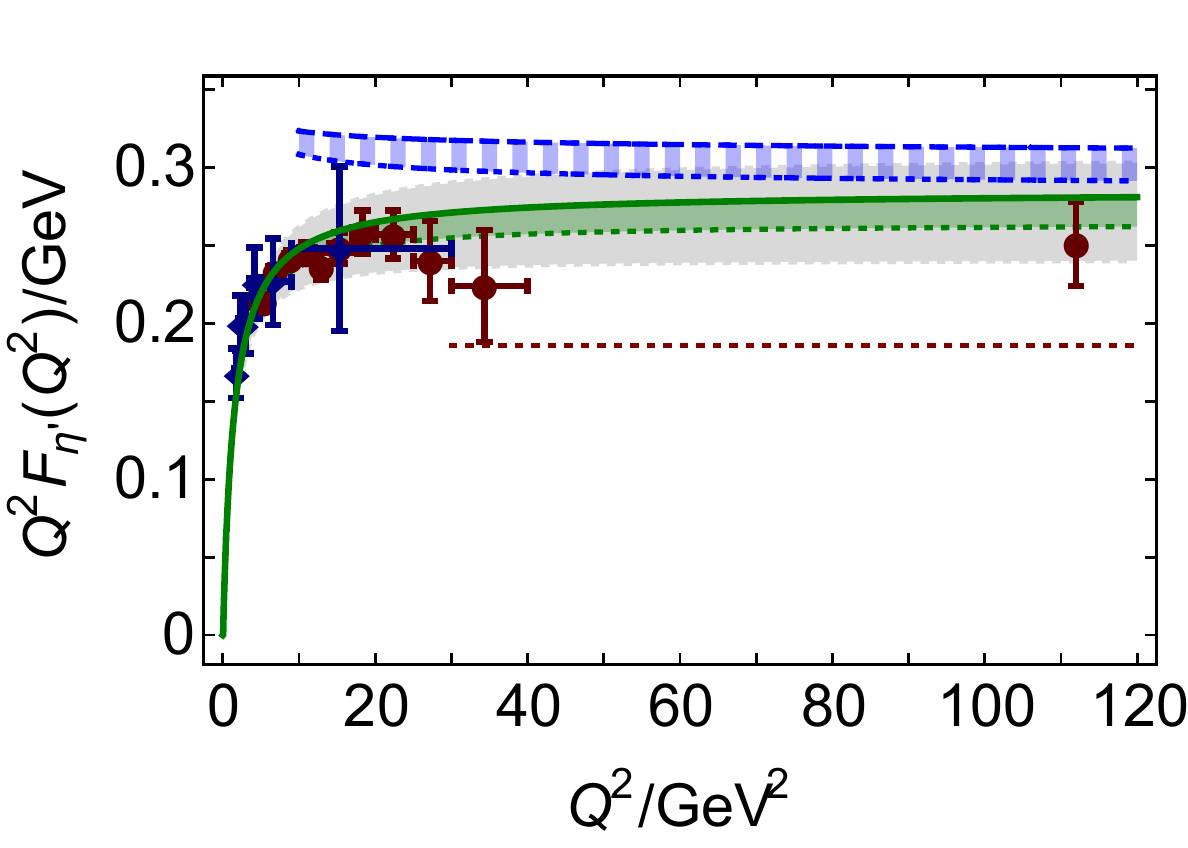}
\end{center}
\caption{\label{TFFslargewQ2} $\gamma^\ast \gamma \to \eta, \eta^\prime$ transition form factors: left panel, $\eta$; right panel, $\eta^\prime$.
Curves: solid (green), our prediction with complete evolution, with the shaded (green) band indicating an uncertainty in our prediction owing to variations in the value of $f_\eta^s$; dashed (blue), asymptotic limit, with the banded-shading (blue) region indicating the impact of uncertainty in $f_\eta^s$ on the asymptotic behaviour. The broader, shaded (grey) band combines this with the uncertainty owing to omission of  $\varphi_M^0$-$\varphi_M^g$ mixing ($\varphi_M^0$ is the flavor singlet distribution amplitude and $\varphi_M^g$ is the two-gluon distribution amplitude). In both panels, the dotted (red) curve is the $\pi^0$ asymptotic limit, $2f_\pi$. Experimental data are\cite{BABAR:2011ad}: diamonds (blue) ``CLEO''; circles (red) ``BaBar''.
}

\end{figure}

The transition form factors $F_{\eta,\eta^\prime}(Q^2)$ obtained within the framework sketched above, are shown in Fig.~\ref{TFFslargewQ2} and compared with the asymptotic value of the $\pi^0$ form factor, $2 f_\pi = 0.186\,$GeV, drawn as the dotted (red) curve in both panels. There are marked similarities between $F_{\eta}(Q^2)$ and the $\pi^0$ transition form factor (see, for example, Ref.~\citenum{Raya:2015gva}, Fig.~2); furthermore, our full $F_{\eta,\eta^\prime}(Q^2)$ result (solid, green curve) agree well with existing data.  Finally, at asymptotically large momentum transfers, \emph{i.e}.\ on $\tau^2 := \Lambda_{\rm QCD}^2/Q^2 \simeq 0$, our full results meet the asymptotic trajectory (dashed blue curve).

The scope of the analysis sketched herein could also be extended to include the doubly off-shell transition form factors, for which the first data $\gamma^\ast(k_1) \gamma^\ast(k_2) \to \eta^\prime$ now exist \cite{BaBar:2018zpn}.\\

\noindent{\em Acknowledgments:} Work partially supported by Jiangsu Province \emph{Hundred Talents Plan for Professionals}.

\end{document}